\documentclass{elsart}

\usepackage{amsmath,amssymb,mathrsfs}

\newcommand{\cc}{\:\text{\sc c}}
\newcommand{\Lag}{\mathscr{L}}
\newcommand{\wt}[1]{\widetilde{#1}}
\newcommand{\hc}{\ensuremath{\mathrm{h.c.}\,}}

\DeclareMathOperator{\imag}{\mathcal{I}\mathit{m}}

\begin{document}

{\normalsize \hfill hep-ph/0604110\mbox{}}\\[2mm]

\begin{frontmatter}

\title{Spontaneous CP Violation in a SUSY Model with a complex CKM}

\author[a,b]{G.C.~Branco}\ead{gbranco@cftp.ist.utl.pt}, 
\author[a]{D.~Emmanuel-Costa}\ead{david.costa@ist.utl.pt}\phantom{,} and  
\author[a]{J.C.~Rom\~ao}\ead{jorge.romao@ist.utl.pt}

\address[a]{\small
  Departamento de F\'{\i}sica and Centro de F\'{\i}sica Te\'orica de 
Part\'{\i}culas (CFTP),
 Instituto Superior T\'ecnico, Av. Rovisco Pais, P-1049-001 Lisboa, Portugal}

\address[b]{\small
Physik-Department, Technische Universit\"at M\"unchen,
James-Franck-Strasse, D-85748 Garching, Germany}


\begin{abstract}
  It is pointed out that the recent measurement of the angle $\gamma$
  of the unitarity triangle, providing irrefutable evidence for a
  complex Cabibbo-Kobayashi-Maskawa (CKM) matrix, presents a great
  challenge for supersymmetric models with spontaneous CP violation.
  We construct a new minimal extension of the minimal supersymmetric
  standard model (MSSM), with spontaneous CP breaking, which leads to
  a complex CKM matrix, thus conforming to present experimental
  data. This is achieved through the introduction of two singlet
  chiral superfields and a vector-like quark chiral superfield which
  mixes with the standard quarks. A $Z_3$ symmetry is introduced in
  order to have a potential solution to the strong CP problem.
\end{abstract}

\end{frontmatter}

\section{Introduction}

Four decades after the experimental discovery of CP violation, the
origin of CP breaking remains one of the fundamental open questions
in particle physics. There are two basic scenarios for CP violation
in the framework of gauge theories, namely explicit CP breaking at
the Lagrangian level and spontaneous CP breaking by the vacuum
(SCPV) \cite{Lee:1973iz,Branco:1979pv}. In the Standard Model
(SM), one assumes that CP is explicitly broken at the Lagrangian
level through the introduction of complex Yukawa couplings leading to
CP violation in the charged-weak interactions, parametrised by a
complex Cabibbo-Kobayashi-Maskawa (CKM) matrix \cite{Cabibbo:1963yz}.

In supersymmetric (SUSY) extensions of the SM there are additional
sources of explicit CP violation, arising from complex soft SUSY
breaking terms as well as from the complex SUSY conserving $\mu$
parameter. The SUSY phases also generate large contributions to the
electric dipole moments (EDMs) of the electron, neutron and mercury
atom. The non-observation of the EDMs imposes strong constraints on
the SUSY phases, forcing them to be very small.  This is the so-called
SUSY CP problem and many solutions have been proposed to overcome it
\cite{abel:2001vy,khalil:2002qp}.

SCPV is an attractive approach to the SUSY CP problem, since all the
couplings of the Lagrangian are real, due to the imposition of CP
invariance at the Lagrangian level. The only source of CP violation are
the vacuum phases~\cite{Lee:1973iz,Branco:1979pv}. SCPV also
provides an appealing solution to the strong CP problem, since in this
case one may have naturally vanishing $\overline{\Theta}$ at tree
level and calculable at higher orders~\cite{Peccei:1988ci,mohapatra:1978fy,georgi:1978xz,barr:1979as,nelson:1983zb,barr:1984qx}.

In this paper, we construct a minimal supersymmetric extension of the
SM, with spontaneous CP breaking, which is in agreement with all the
present data, provided by both BaBar and Belle collaborations
\cite{Aubert:2004qm,Chao:2004mn}. The question of compatibility of
spontaneous CP breaking in supersymmetric (SUSY) models with recent CP
data is highly non-trivial, for the following reason.  One of the
salient features of most SUSY models with spontaneous CP breaking is
the fact that they lead to a real CKM matrix, since the phases in
quark mass matrices can be removed by rephasing of right-handed quark
fields, in a manner entirely analogous to the one encountered in
models with natural flavour conservation in the Higgs sector
\cite{Branco:1979pv}. Until recently, SUSY models with spontaneous CP
breaking and a real CKM matrix \cite{Branco:2000dq,hugonie:2003yu}
could be viable since both $\varepsilon_K$ and $a_{J/\psi\:K_S}$ could
be generated by supersymmetric contributions to $K^0-\overline{K^0}$
and $B_d-\overline{B_d}$ mixing, respectively
\cite{Branco:1995cj}. The novel experimental input is the recent
measurement of the angle $\gamma$ of the unitarity
triangle~\cite{Aubert:2004qm,Chao:2004mn}. A recent analysis
\cite{Botella:2005fc} of the present experimental data provides clear
evidence for a complex CKM matrix even allowing for the presence of
new physics contributions to $\varepsilon_K$, $a_{J/\psi\:K_S}$,
$\Delta M_{B_d}$ and $\Delta M_{B_s}$. These experimental findings
lead to the question of whether it is possible to have a SUSY
extension of the SM with spontaneous CP violation and a complex CKM
matrix. In this paper we will show that the answer to the above
question is in the affirmative. Indeed we construct a minimal
supersymmetric extension of the SM (MSSM) with spontaneous CP breaking
which leads to a complex CKM matrix, thus avoiding conflict with
recent data. The crucial point is the introduction of at least one
$Q=-1/3$ vector-like quark which mixes with standard quarks and leads
to a non-trivial phase in the $3\times3$ effective CKM matrix
connecting standard quarks. The existence of such matter states
naturally arises in various extensions of the SM, such as in $E_6$
grand unified theories as well as in models with extra-dimensions,
including some of the superstring inspired models.

The model presented here has another advantage with respect to
previously suggested SUSY models with SCPV. This stems from the fact
that physical phases only arise in the vacuum expectation values
(VEVs) of Higgs singlet phases. As these singlets only couple to the
quarks and squarks, they will not show up in the neutralino and
chargino sectors, making the solution of the SUSY CP problem easier in
this context.

\section{The Model}

We consider a simple extension of the MSSM where in addition to the
usual superfield content, we introduce an isosinglet vector-like down
quark, $D_4,\,D_4^{\cc}$, and in the Higgs sector, two singlet chiral
superfields $S_1,\,S_2$ which have, under $SU(3)_{\text{\sc C}}\times
SU(2)_{\text{\sc L}}\times U(1)_{\text{\sc Y}}$, the following quantum
numbers
\begin{equation}
  D_4:\,\left(\mathbf{3},\mathbf{1}\right)_{-1/3}\,,\qquad
  D_4^{\cc}:\,\left(\mathbf{\bar3},\mathbf{1}\right)_{1/3}\,,\qquad
  S_1,S_2:\,\left(\mathbf{1},\mathbf{1}\right)_{0}\,.
\end{equation}
We will show that the extended Higgs structure can spontaneously break
CP through phases in the VEVs of the singlet scalar components
$s_1,s_2$. Furthermore, these phases induce a non-vanishing phase in
the CKM matrix, which is not suppressed by the small ratios
$\left|\langle h_{u,d}\rangle/\langle s_{1,2}\rangle\right|\ll 1$. We
emphasise that small values of the later ratios are crucial in order
to naturally suppress flavour-changing neutral currents (FCNC), but
also the one-loop finite contributions to the parameter
$\overline{\Theta}$ associated with strong CP violation
\cite{Bento:1991ez}.

Since we want to achieve spontaneous CP breaking, we impose CP
invariance at the Lagrangian level, which implies that all the
parameters appearing in the superpotential and in the soft breaking
terms are real. Moreover, we introduce a discrete $Z_3$ symmetry,
under which the standard superfields transform trivially, while the
new chiral superfields have the following $Z_3$ assignment
\begin{equation}
\left\{D_4^{\cc},\,S_1,\,S_2\right\}\,\to\, e^{\frac{2\,\pi}3\:i}
\,,\qquad D_4\,\to\, e^{\frac{4\,\pi}3\:i}\,.
\end{equation}
The r\^ole of the $Z_3$ symmetry is to forbid quark bare mass terms of
the form $\bar{d} _{i\,\text{\sc L}} d_{4\,\text{\sc R}}$, thus
leading to the vanishing of $\overline{\Theta}$ at the tree level. The most
general renormalisable gauge invariant superpotential which is
compatible with R-parity invariance and the $Z_3$ symmetry reads as
\begin{equation}
\label{eq:W}
\begin{aligned}
W=& \,\varepsilon_{ab}\left(\,Y_u^{\:ij}\,Q^a_i\,H^b_u\,U_j^{\cc}\,+\,
Y_d^{\:ij}\,Q^b_i\,H^a_d\,D_j^{\cc}
\,+\,Y_e^{\:ij}\,L^b_i\,H^a_d\,E_j^{\cc}\,+\,\mu\,H^a_u\,H^b_d \,\right)\\[2mm]
&\,+\,f^i\,S_1\,D_4\,D^{\cc}_i\,+\,
g^i\,S_2\,D_4\,D^{\cc}_i\,+\,M_D\,D_4\,D^{\cc}_4\\[2mm]
&\,+\,\lambda_{11}\,S^3_1\,+\,\lambda_{12}\,S^2_1\,S_2\,+\,\lambda_{21}\,S_1\,S^2_2
\,+\,\lambda_{22}\,S^3_2\ ,
\end{aligned}
\end{equation}
where the indices $a,b\,=\,1,2$ are $SU(2)$ indices and $\varepsilon$
is a antisymmetric $2\times2$ matrix, with
$\varepsilon_{12}\,=\,1$. We have assumed that the singlets
superfields $S_1,\:S_2$ are even and the isosinglet superfields
$D_4,\,D_4^{\cc}$ are odd under the extended R-parity. The matrices
$Y_u\,,Y_d\,,Y_e$, the vectors $f\,,g$ and $M_D$ generate the masses
of quarks (including the vector-like one) and leptons.
 
In addition to the superpotential given by Eq.(\ref{eq:W}), we have to
specify the explicit soft-breaking terms, which read as
\begin{equation}
\begin{split}
  -&{\Lag}_{\rm soft}
  =\,(M_Q^2)_{ij}\,\wt{q}^{a\:\ast}_i\,\wt{q}^a_j
    \,+\,(M_L^2)_{ij}\,\wt{l}^{a\:\ast}_i\,\wt{l}^a_j
    \,+\,(M_U^2)_{ij}\,\wt{u}^{\cc}_i\,\wt{u}^{\cc\:\:\ast}_j
    \,+\,(M_D^2)_{\alpha\beta}\,\wt{d}^{\cc}_\alpha\,\wt{d}^{\cc\:\ast}_\beta
    \\[2mm]
    &\,+\,(M_E^2)_{ij}\,\wt{e}^{\cc}_i\,\wt{e}^{\cc\:\ast}_j
    \,+\,M_{D_4}^2\,\wt{d}_4\,\wt{d}^{\:\ast}_4 
    +\left(\,M^2_{H_d}\,h_d^{a\:\ast}\,h^{a}_d
    +  \,M^2_{H_u}\,h_u^{a\:\ast}\,h^a_u \right)
    \\[2mm]
    &\,+  \,M^2_{S_1}\,s_1^{\ast}\,s_1
    +  \,M^2_{S_2}\,s_2^{\ast}\,s_2\,+\,
    \varepsilon_{ab}\left(
    A_u^{ij}\,Y_u^{ij}\,\wt{q}^{\,a}_i\,\wt{u}^{\cc}_j\,h^{\,b}_u\,+\,
    A_d^{ij}\,Y_d^{ij}\,\wt{q}^{\,b}_i\,\wt{d}^{\cc}_j\,h^{\,a}_d\,+\right.\\[2mm] 
    &\left.+\,A_e^{ij}\,Y_e^{ij}\,\wt{l}^{\,b}_j\,\wt{e}^{\cc}_i\,h^{\,a}_d\,+\,
    B\,\mu\,h_u^a\,h_d^b
    \,+\,\hc \right)  \\[2mm]        
    &+\,\left(A_{f^i}\,f^i\,s_1\,\wt{d}_4\,\wt{d}_i^{\cc}
    \,+\,A_{g^i}\,g^i\,s_2\,\wt{d}_4\,\wt{d}_i^{\cc}
    \,+\,\hc\right)\\[2mm]
    &+\,\left(A_{\lambda_{11}}\,\lambda_{11}\,s^3_1\,+\,
    A_{\lambda_{12}}\,\lambda_{12}\,s_1^2\,s_2\,+\,
    A_{\lambda_{21}}\,\lambda_{21}\,s_1\,s_2^2\,+\,
    A_{\lambda_{22}}\,\lambda_{22}\,s^3_2\,+\,\hc\right)\\[2mm]
    &- \frac12\left(M_1\,\wt{B}\wt{B}
    +M_2\,\wt{W}\wt{W}+M_3\,\wt{G}\wt{G} + \hc \right)\ .
\end{split}
\label{MSSMsoft}
\end{equation}
In the above equation, we take the soft-SUSY breaking coefficients
$M^2_{Q,\:\dots}$, $B$, $A_{u,\:\dots}$ and $M_{1,2,3}$ as free
parameters at the weak scale, and choose them real in order to respect
CP invariance of the Lagrangian.

\section{Analysis and results}
\subsection{Minimisation of the neutral scalar potential}

After spontaneous electroweak symmetry breaking, in the neutral
Higgs sector, we have verified that there is a region of parameter
space where minimum of the Higgs potential is at:
\begin{equation}
\label{eq:vevs}
\langle h_d\rangle=\begin{pmatrix}v_d\\0\end{pmatrix}\ ,\quad
\langle h_u\rangle=\begin{pmatrix}0\\v_u\,e^{i\,\theta}\end{pmatrix}\ ,\quad
\langle s_1\rangle=V_1\,e^{i\,\phi_1}\ ,\quad
\langle s_2\rangle=V_2\,e^{i\,\phi_2}\ , \quad 
\end{equation}
where $v_d$, $v_u$, $V_1$ are real, positive and $\theta$, $\phi_1$
and $\phi_2$ are physically meaningful phases. It will be shown that
the minimum of the potential is at $\theta=0$. On the contrary, the
phases $\phi_1,\,\phi_2$ are in general non-vanishing and lead to
SCPV. Furthermore, we will show that these phases lead to an effective
down quark mass matrix capable of generating a CP violating phase in
the $3\times3$ sector of the CKM matrix.

From the superpotential and soft SUSY breaking terms,
Eq.(\ref{eq:W}) and Eq.(\ref{MSSMsoft}), we derive the following
CP-invariant neutral scalar potential:
\begin{equation}
\label{eq:1}
\begin{aligned}
&V_{\text{\tiny neutral}}=V^{\text{\tiny MSSM}}_{\text{\tiny neutral}}\,+\,
M^2_{S_1}\, V_1^2 + M^2_{S_2}\, V_2^2 \\[+2mm]
&
+ \left(9\,\lambda_{11}^2+1\,\lambda_{12}^2\right)\,V_1^4 
+ \,4\:\left(\lambda_{12}^2+\lambda_{21}^2\right)\,V_1^2\,V_2^2 
+ \left(9 \, \lambda_{22}^2+1\, \lambda_{21}^2\right)\, V_2^4 \\[+2mm]
&   
+ \, 4\:V_1\, V_2\, (3\, \lambda_{11}\, \lambda_{12}\, V_1^2 
+ \lambda_{21}\, (3\, \lambda_{22}\, V_2^2 + \lambda_{12}\, (V_1^2 + V_2^2)))\, 
\cos{(\phi_1 - \phi_2)}  \\[+2mm]
&
+ 6 \, \lambda_{11}\, \lambda_{21}\, V_1^2\, V_2^2\, \cos{(2\, (\phi_1
  - \phi_2))} + 6 \, \lambda_{12}\, \lambda_{22}\, V_1^2\, V_2^2\, \cos{(2\, (\phi_1 - \phi_2))} \\[+2mm]
&
+ 2\, A_{\lambda_{12}}\,\lambda_{12}\,
 V_1^2\, V_2\, \cos{(2\, \phi_1 + \phi_2)}
 \,+\, 2\: A_{\lambda_{21}}\, \lambda_{21}\, 
V_1\, V_2^2\, \cos{(\phi_1 + 2\, \phi_2)}\\[+2mm]
&
+\, 2\:A_{\lambda_{11}}\,\lambda_{11}\,V_1^3\,\cos{(3\,\phi_1)}
+\, 2\:A_{\lambda_{22}}\,\lambda_{22}\,V_2^3\,\cos{(3\,\phi_2)}\,.
\end{aligned}
\end{equation}
where the MSSM part is
\begin{equation}
\begin{aligned}
  \label{eq:MSSMpot}
    V^{\text{\tiny MSSM}}_{\text{\tiny neutral}}&=\,
\left(M^2_{H_d} + \mu^2 \right) v_d^2 + \left(M^2_{H_u}
+ \mu^2 \right) v_u^2 \,+\,
\frac{1}{8}\left(g^2+g'^2\right)
\left( v_u^2-v_d^2 \right)^2\\[+2.5mm] 
&\,-\,2\: B \mu\:v_u v_d \cos\theta\,.
 \end{aligned}
\end{equation}
 From Eq.~(\ref{eq:MSSMpot}) it is clear that the minimisation
 equations will require $\theta=0$, for $\beta\mu$ positive. Note that
 $\beta\mu$ positive is required, as in the MSSM, in order to obtain
 positive squared masses for the physical CP-odd state contained in
 $h_u,\,h_d$. This is to be expected as the singlets do not mix with
 the MSSM doublets and it is well known that the MSSM does not to lead
 spontaneous CP violation~\cite{Branco:1979pv,Romao:1986jy}. Setting
 $\theta=0$ we still have six minimisation equations for the four VEVs
 and two phases. The minimisation of this potential is quite involved.
 We have done it numerically, following the procedure of
 Ref.~\cite{Romao:1992vu}. To explain it briefly, we write the
 minimisation equations in the form:
\begin{eqnarray}
  \label{eq:2}
  \frac{\partial V_{\text{\tiny neutral}}}{\partial v_d} &=&
2\:\left( M^2_{H_d} + \mu^2\right) v_d - \frac{g^2+
  g'^2}{2}\left(v_u^2-v_d^2\right) v_d -2\:B \mu v_u\,,
\nonumber\\[+2mm]
  \frac{\partial V_{\text{\tiny neutral}}}{\partial v_u} &=&
2\:\left( M^2_{H_u} + \mu^2\right) v_u + \frac{g^2+
  g'^2}{2}\left(v_u^2-v_d^2\right) v_u -2\:B \mu v_d\,,
\nonumber\\[+2mm]
  \frac{\partial V_{\text{\tiny neutral}}}{\partial V_1} &=&
2\:M^2_{S_1} V_1 + F_1(\lambda_{ij},A_{\lambda_{ij}},V_i,\phi_i)\,,\\[+2mm]
  \frac{\partial V_{\text{\tiny neutral}}}{\partial V_2} &=&
2\:M^2_{S_2} V_1 + F_2(\lambda_{ij},A_{\lambda_{ij}},V_i,\phi_i)\,,
\nonumber\\[+2mm]
  \frac{\partial V_{\text{\tiny neutral}}}{\partial \phi_1} &=&
6\:A_{\lambda_{11}} \lambda_{11} V_1^3 \sin\phi_1
+ F_3(\lambda_{ij},A_{\lambda_{ij}},V_i,\phi_i)\,,\nonumber\\[+2mm]
  \frac{\partial V_{\text{\tiny neutral}}}{\partial \phi_2} &=&
6\:A_{\lambda_{22}} \lambda_{22} V_2^3 \sin\phi_2
+ F_4(\lambda_{ij},A_{\lambda_{ij}},V_i,\phi_i)\nonumber\,.
\end{eqnarray}

These equations can be analytically solved for the quadratic soft
masses~\cite{Romao:1992vu} and for the $A_{\lambda_{11}}$ and
$A_{\lambda_{22}}$ parameters~\cite{hugonie:2003yu}. The procedure is
therefore to scan over all the VEVs, phases and remaining parameters,
and obtain $M^2_{H_d}$, $M^2_{H_u}$, $M^2_{S_1}$, $M^2_{S_2}$,
$A_{\lambda_{11}}$, $A_{\lambda_{22}}$ from Eq.~(\ref{eq:2}). We are
then certain that the extrema equations are satisfied. To assure that
we are at a minimum we compute the eigenvalues of the neutral Higgs
boson mass matrices and require that they are all positive, except for
the Goldstone bosons that, of course, should remain massless. If this
is verified we are at a minimum. Finally we check if this minimum is
deeper than the minimum that does not violate CP. Following this
procedure we easily obtained a very large number of good solutions,
which are true minima of the potential and do violate CP. Next we
analyse the question of CP violation in the quark sector.
 
\subsection{The quark mass matrices}

Upon electroweak gauge symmetry breaking, quark mass matrices are
generated through the Yukawa terms, $Y_u\,,Y_d\,,f\,,g$, and the mass
term $M_D$. Note that $Y_u\,,Y_d$ and $M_D$ are real. The couplings $f$
and $g$ are also real, but the mass terms generated by them are in
general complex due to the phases $\phi_i\equiv\arg
\langle s_i\rangle$. Thus, the quark mass matrix are
encoded in the following Lagrangian:
\begin{equation}
-\Lag_{\text{\sc mass}}\,=\,u^T_i\,(m_u)_{ij}\,u^{\cc}_j\,+\, 
d^{\:T}_{\alpha}\,(M_{d})_{\alpha\beta}\, d^{\cc}_{\beta} 
\,+\,\hc\ ,
\end{equation}
where $i,j=1,2,3$, $\alpha=1,2,3,4$, and the matrices $m_u$ and $M_d$
are given
\begin{equation}
\label{Md}
m_u=v_u\,Y_u\,,
\qquad
M_d\,=\,
\begin{pmatrix}
  m_d && 0 \\[1mm]
  M && M_D  
\end{pmatrix}\,, 
\end{equation}
with $m_d=v_d\,Y_d$, and $M_i= f_i\,V_1\,e^{i\phi_1}\,+\,
g_i\,V_2\,e^{i\phi_2}$. In order to simplify our analysis, and without
loss of generality, we perform an orthogonal weak basis transformation
that leaves the matrix $Y_u$ in a diagonal form. The complex elements
$M_i$ can be written in terms of moduli and phases as
\begin{equation} 
  M=\left(|M_1|\,e^{i\,\varphi_1},\,|M_2|\,e^{i\,\varphi_2},\,
|M_3|\,e^{i\,\varphi_3}\,\right)\,,
\label{eq:M}
\end{equation}
where these moduli and phases are given by
\begin{equation}
\label{eq:Mi}
  |M_i|^2\,=\,f_i^2V^2_1+g_i^2V^2_2+2f_ig_iV_1V_2\cos(\phi_1-\phi_2)\,,
\end{equation}
and 
\begin{equation}
\label{eq:phii}
\tan\varphi_i\,=\,\frac{f_iV_1\sin\phi_1+g_iV_2\sin\phi_2}{
  f_iV_1\cos\phi_1+g_iV_2\cos\phi_2}\,,
\end{equation}
respectively. Moreover, the phase $\varphi_1$ can be set to zero,
without loss of generality, by re-definition of the vector-like quark
fields. The phases $\varphi_2$ and $\varphi_3$ are the only source of
CP violating that appear encoded on the $3\times4$ CKM matrix. In
what follows, we assume $m_d$ to be a $3\times3$ real symmetric
matrix. Since the mass terms $M\,,M_D$ are $\Delta I=0$, they can be
much larger than $v_u$ and $v_d$.

As we have mentioned in the introduction, the strong CP problem is
automatically solved at tree level. The parameter $\overline{\Theta}$
associated with strong CP problem can be written as
$\overline{\Theta}=\Theta_{\text{\tiny\sc
QCD}}\,+\,\Theta_{\text{\tiny \sc QFD}}$, where $\Theta_{\text{\tiny
\sc QCD}}$ is the coefficient of $g_s^2 F\tilde{F}/32\pi^2$ which
vanishes since CP symmetry is imposed at the Lagrangian level. The
parameter $\Theta_{\text{\tiny \sc QFD}}$, at tree level, is given by
\begin{equation}
\Theta_{\text{\tiny \sc QFD}}=\arg\left[\det(m_u)\,\det(M_d)\right]\,,
\end{equation}
 since the matrix $m_u$ and the determinant of the matrix 
$M_d$ from Eq.(\ref{Md}) are real,
\begin{equation}
\label{eq.det}
\det\left(M_d\right)\,=\,M_D\,\det\left(m_d\right)\,,
\end{equation}
it follows that  
\begin{equation}
\label{eq:scpp}
\overline\Theta_{\:\text{\tiny tree}}\,=\,0\,,
\end{equation}
thus providing a strong CP problem solution  of the
Barr-Nelson type \cite{nelson:1983zb,barr:1984qx,barr:1988wk}. Note that the
determinant of Eq.(\ref{eq.det}) is real due to the $Z_3$ symmetry,
which forbids terms like $Q_i\, H_d\, D_4^{\cc}$.

In order to determine the left-handed unitary rotations which encodes
the quark-isosinglet mixings and the CP violation information one has
to diagonalise the hermitian matrix
$H=M_d\,M^{\dagger}_d$ ($4\times4$) written as
\begin{equation}
\label{eq:h}
H\,=\begin{pmatrix}m_d^2 && m_d\,M^{\dagger}\\M\,m_d && \overline{M}^2
\end{pmatrix}\,,
\end{equation}
where $\overline{M}^{\:2}=|M_1|^2+|M_2|^2+|M_3|^2+M_D^2$. The matrix
$H$ can be diagonalised by the following unitary matrix
\cite{Bento:1991ez}
\begin{equation}
U\,=\,
\begin{pmatrix}
K & R \\
S & T
\end{pmatrix}\,,
\label{eq:diagU}
\end{equation}
leading to:
\begin{equation}
U^{\dagger}\,H\,U\,=\,\begin{pmatrix}
\mathcal{D}^2_d & 0 \\
0 & \widetilde{M}^2_D
\end{pmatrix}\,,
\label{eq:diageq}
\end{equation}
where $\mathcal{D}^2_d$ is diagonal formed with the light down quark
squared masses. From Eq.(\ref{eq:diagU}) and Eq.(\ref{eq:diageq}) one
derives:
\begin{equation}
\label{eq:exeff}
\widetilde{H}\,\equiv\,K\,\mathcal{D}^2_d\,K^{-1}\,=\,
m_d\,m_d^{\dagger}-
  \frac{m_d\,M^{\dagger}M\,m_d^{\dagger}}{\overline{M}^2}\,
\left(\mathbf{1}-\frac{\widetilde{H}}{\overline{M}^2}\right)^{-1}\,.
\end{equation}
An effective $3\times3$ light down quark mass matrix can be easily
obtained in the limit where $K$ is an unitary matrix and
$\widetilde{H}\ll\overline{M}^2$
\cite{Bento:1991ez,Grimus:2000vj}. One obtains:
\begin{equation}
\label{eq:meff}
m_{\text{eff}}m^{\dagger}_{\text{eff}}\,\approx
m_d\:\left( \mathbf{1}- \frac{M^{\dagger}M}{\overline{M}^{\:2}}\right)\:m_d\,,
\end{equation}
while the mass of the down-type quark isosinglet is given in an
excellent approximation by
\begin{equation}
\label{eq:wtM}
  \widetilde M_D\approx \overline M\,.
\end{equation}
Note that the matrix $K$ is almost unitary since from
Eq.(\ref{eq:diageq}) and exact unitarity of $U$, it follows that
\begin{equation}
K^{\dagger}K\,=\,1-S^{\dagger}S\,,
\end{equation}
with $S\,\approx\,\mathcal{O}(m_d\,/\,\overline M)$
\cite{Bento:1991ez}. Although, the matrix $m_d$ is real, the phases in
$M^{\dagger}M$ in Eq.(\ref{eq:meff}) are sufficient to generate a
complex $K\:$. The full CKM matrix, which now has dimension
$3\times4$, comes directly from the matrix $U$ and is given by:
\begin{equation}
\label{eq:ckm34}
  (V_{\text{\tiny\sc CKM}})_{i\alpha}=
\begin{pmatrix} K && R\end{pmatrix}_{i\alpha},
\end{equation}
which contributes to the electroweak Lagrangian as:
\begin{equation}
\begin{split}
\label{wesnew}
{\Lag}_{\text{\sc W,Z}} & =\,  -\frac{g}{\sqrt 2} \left ( \bar
u_{L\:i} \gamma^{\mu} \,\left(V_{\text{\tiny\sc CKM}}\right)_{i\,\alpha}
d_{L\:\alpha}\, W_\mu^{+} + {\mathrm h.c.}
\right )\,+\\ & -\frac{g}{2\cos\,\theta_{\mathrm W}} \left ( \bar
u_{L\:i} \gamma^\mu u_{L\:i}- \bar d_{L\:\alpha} \gamma^\mu X_{\alpha\beta}\,
d_{L\:\beta} - 2\sin^2\theta_{\mathrm W}\, J_{\mathrm EM}^\mu \right )\:Z_{\mu}\,,
\end{split}
\end{equation}
where 
\begin{equation}
X_{\alpha\beta}\equiv 
\left(V^{\dagger}_{\text{\tiny\sc CKM}}\,
V_{\text{\tiny\sc CKM}}\right)_{\alpha\beta}
\,=\,\delta_{\alpha\beta}-\begin{pmatrix}S^{\dagger}S&&S^{\dagger}T\\[1mm]
T^{\dagger}S && T^{\dagger}T\end{pmatrix}_{\alpha\beta},
\end{equation}
and the electromagnetic current, $J_{\mathrm EM}^\mu$, is given by
\begin{equation}
  J_{\mathrm EM}^\mu =
  \frac{2}{3} \bar u_i \gamma^\mu u_i-\frac{1}{3} \bar d_{\alpha} \gamma^\mu
  d_{\alpha}\,.
\end{equation}
The $3\times4$ CKM matrix from Eq.(\ref{eq:ckm34}) encodes three independent
phases which violate CP. However in the limit where one neglects the
small mixings between the physical heavy quark $d_4$ and the light
quarks (\emph{i.e.} the limit where $K$ becomes unitary) only one
phase appears in $K$ \cite{Branco:1986my}.

\subsection{Numerical Example}

In this section we present one concrete numerical example which leads
to values of quark masses and mixings in agreement with
experiments. It is clear from the effective mass matrix presented in
Eq.(\ref{eq:meff}), that the light quarks and their mixing do not
depend on the overall scale, which only rescales the heavy vector-like
down quark mass. We have verified numerically that this feature still
holds, at a very high precision, provided $M_D\gtrsim 1\:\text{ TeV}$,
which enforces the goodness of the effective matrix given in
Eq.(\ref{eq:meff}). We have then parametrised the input values $M_i$
as a function of an overall scale $M_D=1$ TeV. Moreover, we fixed the
down quark masses to the following values:
\begin{equation}
m_d=4.38\text{ MeV}\,,\quad  m_s=94.6\text{ MeV}\,,\quad  
m_b=3.11\text{ GeV}\,.
\end{equation}
We have verified, using Eq.(\ref{eq:exeff}) that there are solutions
to the matrix $m_d$ leading to the effective $3\times3$ CKM matrix $K$
consistent with experimental data. This was done by varying the ratios
$M_i/M_D$ and the others parameters in a consistent way with
Eq.(\ref{eq:exeff}). A particular example is obtained with the
following values for the ratios $M_i/M_D$:
\begin{equation}
\frac{M_1}{M_D}=0.5\,,\quad \frac{M_2}{M_D}=0.8\,, \quad
\frac{M_3}{M_D}=3.8\,, 
\end{equation}
and the  following values for the phases $\varphi_i$ (in radians):
\begin{equation}
\varphi_2=1.1\,,\quad \varphi_3=1.7\,.
\end{equation}
We considered the moduli of matrix $K$ elements as
\begin{equation}
\left|V^{3\times3}_{\text{\tiny CKM}} \right|\,\equiv\,|K|
\,=\,\begin{pmatrix}
  0.97  && 0.22  && 0.0038\\[1mm]
  0.22 && 0.97  && 0.04\\[1mm]
  0.0086 && 0.04  && 0.9992
\end{pmatrix}\,,
\end{equation}
that are within the range allowed by
experiments~\cite{Eidelman:2004wy}. The relevant CP violating weak
basis invariants that encodes phases content in $K$ are
\begin{equation}
J\,\equiv\,\imag\left(K_{12}K^{\ast}_{13}K^{\ast}_{22}K_{23}\right)
\,=\,3.14\times 10^{-5}\,,
\end{equation}
and 
\begin{equation}
\beta\,\equiv\,\arg\left(-\,K_{21}K^{\ast}_{23}K^{\ast}_{31}K_{33}\right)
\,=\,23.8^{\circ}\,,
\end{equation}
which corresponds to:
\begin{equation}
\sin2\beta\,=\,0.739\,,
\end{equation}
in agreement with the most recent experimental
data~\cite{Eidelman:2004wy}. Note that the quantities $J$ and
$\sin2\beta$ are only relevant to describe CP violation in the quark
sector when $K$ is nearly unitarity. The deviation from unitarity of
the matrix $K$ is given by
\begin{equation}
  \mathbf{1}-K^{\dagger}K=S^{\dagger}S\,\simeq\,\mathcal{O}\left(10^{-7}\right)\,,
\end{equation}
which confirms our previous assertions about the almost unitarity of
$K$. Thus, the values of the elements of $m_d$ obtained are:
\begin{equation}
m_d \,=\,\begin{pmatrix}
  0.01  &&  0.025  &&  0.02\\[1mm]
  0.025  &&  0.13  &&  0.47\\[1mm]
  0.02  &&  0.47  &&  9.67
\end{pmatrix}\,.
\end{equation}
Once the matrix $m_d$ is determined, the mass of the heavy vector-like
down quark state can be calculated by using
Eqs.(\ref{eq:h}-\ref{eq:diageq}), giving the following result:
\begin{equation}
 \wt{M}_D\simeq\overline{M}=4.0 \text{ TeV}\,,
\end{equation}
which enforces the excellent approximation given in Eq.(\ref{eq:wtM}).

Finally we have to verify if the inputs are compatible with the
minimisation of the potential. The only parameters that enter the
quark sector and that are relevant for the minimisation of the
potential are the VEVs $V_1$, $V_2$ and the phases $\phi_1$ and
$\phi_2$. As can be seen from Eq.~(\ref{eq:Mi}) and
Eq.~(\ref{eq:phii}) there is a significant freedom in obtaining the
values of $V_i$ and $\phi_i$ from $M_i$ and $\varphi_i$. We have
numerically verified that there is a large number of possible choices
of parameters that give a successful minimisation for a given choice
of $M_i$ and $\varphi_i$, thus implying that our solutions are fully
consistent.

\section{Discussion and Conclusions}

We have presented a new minimal supersymmetric extension of the SM,
where CP is spontaneously broken. The novel feature of the model is
the fact the CKM matrix is non-trivially complex, in contrast with
previous SUSY models with SCPV.

It was emphasised that having a complex CKM matrix is crucial, in view
of the recent measurement of the angle $\gamma$ of the unitarity
triangle. Prior to this important experimental result, it was possible
to have viable SUSY extensions of SM, with SCPV and a real CKM matrix,
where CP violation in the kaon and B-sectors was entirely generated by
new SUSY contributions to $K^0-\overline{K^0}$ and $B-\overline{B}$
mixings. This class of models are no longer viable. In order to
generate a non-trivial CP-violating phase in the CKM matrix, we have
introduced a $Q=-1/3$ heavy isosinglet quark. The mass terms mixing
this heavy quark with the standard quarks are responsible for the
generation of a complex effective $3\times3$ CKM matrix, which in turn
leads to a complex CKM matrix.

Due to the appearance of naturally suppressed FCNC at tree level, in
the down-quark sector, there are in the model new contributions
\cite{Branco:1994jw} to $K^0-\overline{K^0}$, $B_d-\overline{B}_d$ and
$B_s-\overline{B}_s$ mixings. At present, all experimental data on CP
violation as well as on rare $K$ and $B$ decays is in impressive
agreement with the SM. It seems by now clear that the SM and its CKM
mechanism of mixing and CP violation give the dominant contributions
to the physical quantities entering the standard unitarity
triangle. However, there is still room for New Physics, that can even
give a dominant contribution to physical quantities which do not enter
directly in the standard unitarity triangle, an example being the
invariant phase
$\chi\equiv\arg(-V_{ts}V^{\ast}_{tb}V^{\ast}_{cs}V_{cb})$. This phase
has not been measured yet, but it will be measured at LHCb.

We have also emphasised that the class of models presented here, have
the potential of alleviating the so-called SUSY CP-problem. In
particular, we have pointed out that since CP violation arises
entirely from VEVs of the Higgs singlets, they will not show up in the
neutralino and chargino sectors.

A complete analysis of the phenomenological implications of this
model, including the evaluation of EDMs is outside the scope of this
paper and it will be presented elsewhere.

The fact that a minimal realistic extension of the MSSM, capable of
generating spontaneous CP violation, requires the introduction of
vector-like quarks, provides further motivation for the search of
these heavy fermions which arise in a variety of extensions of the SM.

\section*{Acknowledgments}

This work was partially supported by Funda\c{c}\~{a}o para a
Ci\^{e}ncia e a Tecnologia (FCT, Portugal), through the projects
POCTI/FNU/44409/2002, PDCT/FP/ FNU/50250/2003, POCI/FP/63405/2005,
POCI/FP/63415/2005, POCTI/FP /FNU/50167/2003, by the CFTP-FCT UNIT
777, which are partially funded through POCTI (FEDER), and by the
European Commission Human Potential Program RTN network
MRTN-CT-2004-503369.  The work of G.C.B. was supported by the
Alexander von Humboldt Foundation through a Humboldt Research
Award. G.C.B. would like to thank Andrzej J. Buras for the kind
hospitality at TUM.  D.E.C. is supported by \emph{Funda\c{c}\~{a}o
para a Ci\^{e}ncia e a Tecnologia} (FCT, Portugal) under the grants
SFRH/BPD/1598/2000.


\begin{thebibliography}{10}

\bibitem{Lee:1973iz}
  T.~D.~Lee,
  \newblock Phys.\ Rev.\ D {\bf 8} (1973) 1226.

\bibitem{Branco:1979pv}
  G.~C.~Branco,
  \newblock Phys.\ Rev.\ Lett.\  {\bf 44} (1980) 504;
  G.~C.~Branco,
  \newblock Phys.\ Rev.\ D {\bf 22} (1980) 2901;
  G.~C.~Branco, A.~J.~Buras and J.~M.~Gerard,
  \newblock Nucl.\ Phys.\ B {\bf 259} (1985) 306.

\bibitem{Cabibbo:1963yz}
  N.~Cabibbo,
  \newblock Phys.\ Rev.\ Lett.\  {\bf 10} (1963) 531;
  M.~Kobayashi and T.~Maskawa,
  \newblock Prog.\ Theor.\ Phys.\  {\bf 49} (1973) 652.

\bibitem{abel:2001vy}
S.~Abel, S.~Khalil and O.~Lebedev,
\newblock Nucl. Phys. {\bf B606} (2001) 151, [hep-ph/0103320].

\bibitem{khalil:2002qp}
S.~Khalil,
\newblock Int. J. Mod. Phys. {\bf A18} (2003) 1697, [hep-ph/0212050].

\bibitem{Peccei:1988ci} 
  R.~D.~Peccei,
  \newblock Adv.\ Ser.\ Direct.\ High Energy Phys.\  {\bf 3} (1989) 503, 
  in: CP Violation (World Scientific, Singapore);
  J.~E.~Kim,
  \newblock Phys.\ Rept.\  {\bf 150} (1987) 1;
  H.~Y.~Cheng,
  \newblock Phys.\ Rept.\  {\bf 158} (1988) 1.

\bibitem{mohapatra:1978fy}
R.~N. Mohapatra and G.~Senjanovic,
\newblock Phys. Lett. {\bf B79} (1978) 283.

\bibitem{georgi:1978xz}
H.~Georgi,
\newblock Hadronic J. {\bf 1} (1978) 155.

\bibitem{barr:1979as}
S.~M. Barr and P.~Langacker,
\newblock Phys. Rev. Lett. {\bf 42} (1979) 1654.

\bibitem{nelson:1983zb}
A.~E. Nelson,
\newblock Phys. Lett. {\bf B136} (1984) 387.

\bibitem{barr:1984qx}
S.~M. Barr,
\newblock Phys. Rev. Lett. {\bf 53} (1984) 329.

\bibitem{Aubert:2004qm}
  B.~Aubert {\it et al.}  [BaBar Collaboration],
\newblock Phys.\ Rev.\ Lett.\  {\bf 93} (2004) 131801;
 \newblock hep-ex/0408088.

\bibitem{Chao:2004mn}
  Y.~Chao {\it et al.}  [Belle Collaboration],
  \newblock Phys.\ Rev.\ Lett.\  {\bf 93} (2004) 191802;
  K.~Abe {\it et al.}  [Belle Collaboration],
  \newblock hep-ex/0411049.

\bibitem{Branco:2000dq}
  G.~C.~Branco, F.~Kruger, J.~C.~Romao and A.~M.~Teixeira,
  \newblock JHEP {\bf 0107} (2001) 027.

\bibitem{hugonie:2003yu}
  C.~Hugonie, J.~C. Romao and A.~M. Teixeira,
  \newblock JHEP {\bf 0306} (2003) 020.

\bibitem{Branco:1995cj}
  G.~C.~Branco, G.~C.~Cho, Y.~Kizukuri and N.~Oshimo,
  \newblock Nucl.\ Phys.\ B {\bf 449} (1995) 483;
  G.~C.~Branco, G.~C.~Cho, Y.~Kizukuri and N.~Oshimo,
  \newblock Phys.\ Lett.\ B {\bf 337} (1994) 316;
  S.~A.~Abel and J.~M.~Frere,
  \newblock Phys.\ Rev.\ D {\bf 55} (1997) 1623;
  S.~Khalil, T.~Kobayashi and A.~Masiero,
  \newblock Phys.\ Rev.\ D {\bf 60} (1999) 075003;
  D.~A.~Demir, A.~Masiero and O.~Vives,
  \newblock Phys.\ Rev.\ D {\bf 61} (2000) 075009;
  M.~Brhlik, L.~L.~Everett, G.~L.~Kane, S.~F.~King and O.~Lebedev,
  \newblock Phys.\ Rev.\ Lett.\  {\bf 84} (2000) 3041;  
  S.~Baek, J.~H.~Jang, P.~Ko and J.~h.~Park,
  \newblock Phys.\ Rev.\ D {\bf 62} (2000) 117701.

\bibitem{Botella:2005fc}
  F.~J.~Botella, G.~C.~Branco, M.~Nebot and M.~N.~Rebelo,
  Nucl.\ Phys.\ B {\bf 725} (2005) 155.

\bibitem{Bento:1991ez}
  L.~Bento, G.~C.~Branco and P.~A.~Parada,
 \newblock Phys.\ Lett.\ B {\bf 267} (1991) 95;
  L.~Bento and G.~C.~Branco,
 \newblock Phys.\ Lett.\ B {\bf 245} (1990) 599.

\bibitem{Romao:1986jy}
J.~C. Romao,
\newblock Phys. Lett. {\bf B173} (1986) 309.

\bibitem{Romao:1992vu}
J.~C. Romao, C.~A. Santos and J.~W.~F. Valle,
\newblock Phys. Lett. {\bf B288} (1992) 311.

\bibitem{barr:1988wk}
  S.~M. Barr and A.~Masiero,
  \newblock Phys. Rev. {\bf D38} (1988) 366;
  S.~M. Barr and A.~Masiero,
  \newblock Phys. Rev. Lett. {\bf 58} (1987) 187.

\bibitem{Grimus:2000vj}
  W.~Grimus and L.~Lavoura,
  \newblock JHEP {\bf 0011} (2000) 042.

\bibitem{Branco:1986my}
  G.~C.~Branco and L.~Lavoura,
 \newblock Nucl.\ Phys.\ B {\bf 278} (1986) 738.

\bibitem{Eidelman:2004wy}
  S.~Eidelman {\it et al.}  [Particle Data Group],
  \newblock Phys.\ Lett.\ B {\bf 592} (2004) 1.

\bibitem{Branco:1994jw}
  G.~C.~Branco, P.~A.~Parada, T.~Morozumi and M.~N.~Rebelo,
  \newblock Phys.\ Rev.\ D {\bf 48} (1993) 1167.

\end{thebibliography}
\end{document}